\documentclass[twocolumn,secnumarabic,amssymb, nobibnotes, aps, prd]{revtex4-1}
\usepackage{graphicx} 
\usepackage{color}
\draft 
\begin{document}

\preprint{versao 2}

\title{Theoretical study of the hyperfine field at Cu impurities diluted in an iron host}

\author{A. L. de Oliveira}
\email[]{alexandre.oliveira@ifrj.edu.br}

\affiliation{Instituto Federal de Educa\c{c}\~{a}o, Ci\^{e}ncia e Tecnologia do Rio de Janeiro, Campus Nil\'{o}polis, Rua L\'{u}cio Tavares 1045, 26530-060, Nil\'{o}polis, RJ, Brazil}

\author{C. M. Chaves}
\email[]{cmch@cbpf.br}

\affiliation{Centro Brasileiro de Pesquisas F\'{i}sicas, Rua Dr. Xavier Sigaud 150, 22290-180,  Rio~de~Janeiro, RJ, Brazil}

\author{N. A. de Oliveira}

\affiliation{Instituto de F\'{i}sica Armando Dias Tavares, Universidade do Estado do Rio de Janeiro, Rua S\~{a}o Francisco Xavier 524, 20550-013, Rio~de~Janeiro, RJ, Brazil}

\author{A. Troper}

\affiliation{Centro Brasileiro de Pesquisas F\'{i}sicas, Rua Dr. Xavier Sigaud 150, 22290-180,  Rio~de~Janeiro, RJ, Brazil}

\date{\today}

\begin{abstract}
Magnetic hyperfine field at Cu isotopes as impurities in Fe were recently measured at low temperature. A model to explain these experimental results is proposed. The diluted Cu impurities in the ferromagnetic Fe host are described by an extension of the Daniel-Friedel model, including the next neighbor perturbation . In order to account for the available experimental data in Cu isotopes with atomic masses $A$ = 59, 67, 69 and 71 as impurity, we needed to incorporate the Cu anomaly volume in the effective charge to be screened and self-consistent procedures.
\end{abstract}
\maketitle 

\section{Introduction}

Magnetic hyperfine fields at Cu impurities in Fe were recently obtained~\cite{Golovko2011} combining resonance frequencies from experiments involving $\beta$-NMR on oriented nuclei on $^{59}$Cu, $^{69}$Cu, and $^{71}$Cu with magnetic moment values from collinear laser spectroscopy measurements at low temperature. The formation of local magnetic moments in impurities embedded in metallic systems has been the concern of condensed matter theorists since the pioneer work of Friedel in the sixties. Motivated by the mentioned experimental work, a microscopic model to explain  the experimental results for the magnetic hyperfine field is proposed. The diluted Cu impurities in the ferromagnetic transition metal Fe  are described by an extension of the Daniel-Friedel model~\cite{Daniel1963},including the next neighbor perturbation~\cite{Troper2001}. In order to study the different Cu isotopes with $A$ = 59, 67, 69 and 71 as impurity, we had to incorporate the Cu anomaly volume~\cite{Troper2001, Oliveira1998} in the effective charge to be screened (see below). With a Hamiltonian to be presented in the next section and self-consistent procedures, we were able to account for the available experimental data.

\section{The model}

The magnetic moment formation at a Cu impurity diluted in Fe arises from the following effects: i) the charge difference between Fe and the impurity produces an electrostatic  potential which the Fe conduction electron gas will shield;  ii) a magnetic field produced by the \textit{d} Fe band that acts in the impurity through its Fe neighbors~\cite{Troper2001, chaves2011} and also in the whole host. Both contributions act differently on the impurity density of up and down spins thus producing a polarized conduction band and a magnetization $m_0 = n_{0\uparrow}- n_{0\downarrow}$ both in the impurity and in the host as well.

The Hamiltonian that systematizes these effects and allows the calculation of the magnetic moment of the impurity is the following
\begin{equation}
\mathcal{H}=H_{\rm Fe}+V,
\label{eq:ham1}
\end{equation}
where 
\begin{equation}
H_{\rm Fe} = \sum_{ i,\sigma }\varepsilon _{\sigma }^{\rm Fe}c_{i\sigma}^{\dagger}c_{i\sigma }
+\sum_{i,j, \sigma }t_{ij}c_{i\sigma }^{\dagger}c_{j\sigma},
\label{eq:HM}
\end{equation}
defines a pure Fe \textit{s-p} host which consists of a conduction  band polarized by the magnetized \textit{d} host band.  In Eq.~(\ref{eq:HM})  $\varepsilon _{\sigma }^{\rm Fe}$ is the center of the {\it s-p} energy band, now depending on the spin $\sigma $ orientation ($\sigma = \,\uparrow$ or $\downarrow$), $c_{i\sigma}^{\dagger}$ ($c_{i\sigma}$) is the creation (annihilation) operator of conduction electrons at site $i$ with spin $\sigma $ and $t_{ij}$ is the electron hopping energy between neighboring $i$ and $j$ sites.

The second term of Eq.~(\ref{eq:ham1}) is the potential due to the presence of the impurity at site $i=0$,
\begin{equation}
V=\sum_{\sigma}{V_{0\sigma }c_{0\sigma }^{\dagger}c_{0\sigma } }
+\tau\sum_{l\neq 0,\sigma }t_{0l}\left( c_{0\sigma }^{\dagger }c_{l\sigma }+c_{l\sigma }^{\dagger }c_{0\sigma }\right). 
\label{eq:V}
\end{equation}
$V_{0\sigma}=(\varepsilon _{0\sigma }^{\rm Cu}-\varepsilon _{\sigma }^{\rm Fe})$ is a spin dependent local term, $\varepsilon _{0\sigma }^{\rm Cu}$ being the {\it s-p} impurity state energy level.  Also included in $V$ is the change of the nearest-neighbor hopping  due to the breaking of translational invariance by the impurity. The parameter $\tau $ takes into
account the change in the hopping energy associated with the
presence of the impurity~\cite{Acker91b, Oliveira95, Oliveira2003}, $\tau=0$ meaning no disorder in the hopping. 

Using the Dyson equation
\begin{equation}
G_{jl\sigma}(z)=g_{jl\sigma}(z)+g_{j0\sigma}(z)\,V\,G_{0l\sigma}(z),
\end{equation}
the local Green's function $G_{00\sigma}(z)$ due the charge perturbation at the origin, is
\begin{equation}
G_{00\sigma}(z) = \frac{g_{00\sigma }(z)}{\alpha ^{2}-g_{00\sigma }(z)V_{\rm eff}^{\sigma
}(z)},
\end{equation}
where $\alpha = \tau + 1$, $g_{00\sigma }(z)$ is the local Green function for the pure Fe host and
\begin{equation}
V_{{\rm eff}}^{\sigma }(z)=V_{0\sigma }+(\alpha ^{2}-1)(z-\varepsilon^{\rm Fe}_{\sigma } ).
\end{equation}

Assuming that the screening of the charge
difference is made by the {\it s-p} band, the potentials $V_{0\sigma}$ are self-consistently determined in such a way that $\Delta Z_{\sigma }$ gives the total charge difference  $\Delta Z$ between host and impurity
\begin{equation}
\Delta Z=\Delta Z_{\uparrow }+\Delta Z_{\downarrow },
\label{eq:DeltaZ}
\end{equation}
where $\Delta Z_{\sigma }$  is obtained integrating the change in the density of states $\Delta \rho_{\sigma}$,
\begin{equation}
\Delta \rho_{\sigma} =-\frac{1}{\pi}\textrm{Im}\sum_{j}\left(G_{jj\sigma}(z)-g_{jj\sigma}(z)\right)
\end{equation}
from the origin up to the Fermi level $\varepsilon_{\rm F}$. So,
\begin{equation}
\Delta Z_{\sigma }=-\frac{1}{\pi }{\rm Im}\ln \left[ \alpha ^{2}-g_{00\sigma }(\varepsilon_{\rm F})\,V_{\rm eff}^{\sigma
}(\varepsilon_{\rm F})\right].
\label{eq:rsoma}
\end{equation}

Once the potential $V_{0\sigma}$ is self-consistently found, the local {\it s-p} density of states per spin direction at the impurity site are calculated by 
\begin{equation}
\rho _{\sigma }(\varepsilon
)=-\frac{1}{\pi} {\rm Im\,}G_{00\sigma }(z).
\end{equation}
The local {\it s-p} electron occupation number, $n_{0\sigma
}$, is obtained by integrating the corresponding
local density of states up to the Fermi level $\varepsilon _{{\rm F}}$. 

The total magnetic moment ($m_{0}$) at a {\it s-p} impurity, given by 
$
m_{0}=n_{0\uparrow }-n_{0\downarrow },
% \label{eq:mm0}
$
i.e., 
\begin{equation}
{m}_{0}=-\frac{1}{\pi }\sum_{\sigma }\int_{-\infty
}^{\varepsilon _{{\rm F}}}{\rm Im\,}\,\frac{\sigma\; g_{00\sigma }(z)}{\alpha ^{2}-g_{00\sigma }(z)V_{\rm eff}^{\sigma
}(z)}\,{\rm d}z.
  \label{eq:mm0b}
\end{equation}

The total magnetic
hyperfine field at the impurity site is 
\begin{equation}
B_{hf}=A(Z_{{\rm imp}})m_{0},
\label{chf}
\end{equation}
where $A(Z_{{\rm imp}})$ is the Fermi-Segr\`{e} contact coupling parameter.

\section{Discussion and Results}

In order to calculate the local moments and the magnetic hyperfine fields at a Cu impurity diluted in Fe we have to fix some  parameters. Here, we adopt a standard paramagnetic {\it s-p} density of states extracted from first-principles calculations~\cite{Oliveira95}. 

The parameter $\alpha $ which renormalizes the hopping energy, was chosen $\alpha\simeq 1$ given the ratio between the extension of the host and impurity {\it s-p} wave functions.    Keeping fixed these parameters, we self-consistently -- through Eq.~(\ref{eq:rsoma}) -- determine the magnetic moment at the Cu impurity site and its corresponding magnetic hyperfine field. 

Once the Cu isotopes have the same charge, the local hyperfine field should be the same. But the experimental results show this is not the case. In fact the charges are the same but the charge densities are not, because of their different volumes. 
We have followed the procedure of Daniel and Friedel~\cite{Daniel1963, Oliveira1998} and incorporated the respective volume of each Cu isotope in an effective charge $\Delta Z^{\prime}$  to be screened,
\begin{equation}
\Delta Z^{\prime}=\Delta Z-\frac{\delta v^{A}}{v_{\rm Fe}},
\label{eq:dv}
\end{equation}
 where  $\delta v^{A} = v_{\rm Fe} - v^{A}_{\rm Cu}$; $v_{\rm Fe}$ and $v^{A}_{\rm Cu}$ are Fe volume and Cu volume respectively. The values of ${\delta v^{A}}/{v_{\rm Fe}}$ are shown in Table~\ref{tab:deltav}.

\begin{table}[h]
\caption{Relative volume variation for each isotopes.}
\begin{tabular}{c|c|c|c|c}
\hline\hline
 $A$                                                 & 59      &  67       & 69    &  71   \\ \hline
 ${\delta v^{A}}/{v_{\rm Fe}}$    & $\; 0.07465$     & $-0.05082$  &$-0.08219$ &  $-0.11356$  \\
\hline\hline 
\end{tabular}
\label{tab:deltav}
\end{table}

Now the potentials $V_{0\sigma}$ are self-consistently determined in such away to give the effective total charge difference $\Delta Z ^{\prime}$.
 
\begin{figure}[!ht]
\includegraphics [angle=0,width=8cm] {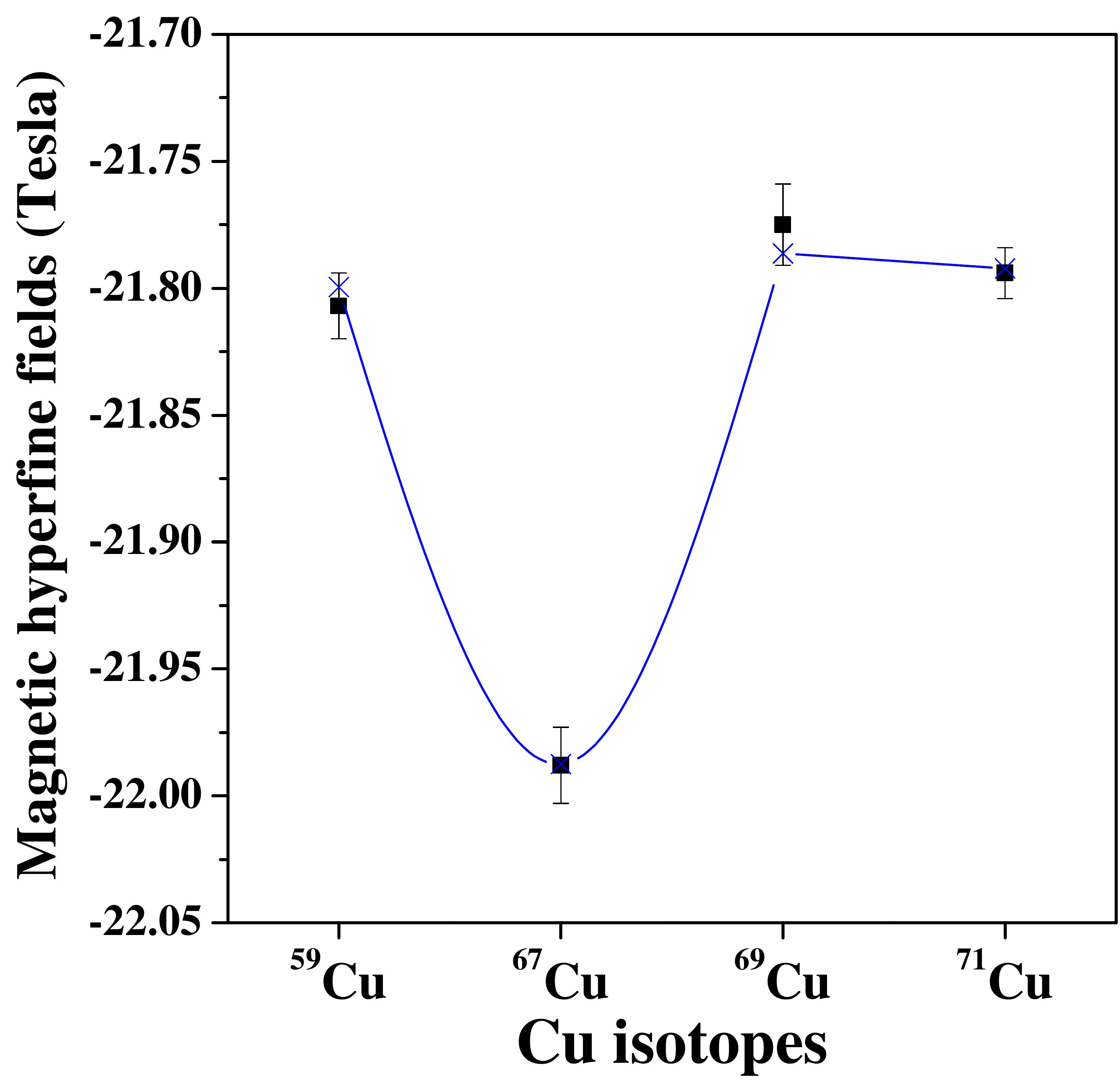}       
\caption{(Color online) Self-consistent calculation of hyperfine fields for Cu isotopes in Fe host incorporating the volume effect (see Eq.~(\ref{eq:dv})). The squares represent the experimental data.} 
\label{fig:Bhf_CuFe}
\end{figure}

The results for the calculated magnetic hyperfine fields (Eq.~(\ref{chf})-(\ref{eq:dv})), shown in Fig.~\ref{fig:Bhf_CuFe}, are in a very good agreement with the experimental results.

Using a very simple microscopic model Hamiltonian, we were able to bring out the isotopic dependence of the hyperfine field at the impurity in Cu\underline{Fe}  therefore extending previous approaches by taking into account the volume of the impurity in the self-consistency procedure. Rather than considering nuclear probes as being punctiform, nuclear experimentalists have now new windows to explore the contribution of the nuclear volume in the interaction between nuclear probes and screening host electrons . 

\section*{Acknowledgments}
We acknowledge the support from the Brazilian agencies CNPq and FAPERJ.
Acker91b, Oliveira95, Oliveira2003

\end{document}